\documentclass[lettersize,journal]{IEEEtran}
\usepackage{amsmath,amsfonts}
\usepackage{algorithmic}
\usepackage{algorithm}
\usepackage{array}
\usepackage{diagbox}
\usepackage{fancyhdr}
\usepackage{textcomp}
\usepackage{stfloats}
\usepackage{url}
\usepackage{tablefootnote}
\usepackage{footnote}
\makesavenoteenv{tabular}
\makesavenoteenv{table}
\usepackage{verbatim}
\usepackage{graphicx}
\usepackage{cite}
\usepackage{xcolor}
\usepackage{tabularray}
\usepackage{wasysym}
\usepackage{algorithmic}
\usepackage{graphicx}
\usepackage{textcomp}
\usepackage{xcolor}
\usepackage{xspace}
\usepackage{subfigure}
\usepackage{multirow}
\usepackage{url}

\usepackage{bbding}
\usepackage{pifont} 
\usepackage{array}
\usepackage{diagbox}
\usepackage{subcaption}
\usepackage{float}
\usepackage[most]{tcolorbox}
\usepackage[export]{adjustbox}
\usepackage{caption}
\usepackage{hyperref}
\usepackage{booktabs,caption}
\usepackage{colortbl}
\usepackage{hhline}
\usepackage[flushleft]{threeparttable}
\usetikzlibrary{automata,positioning}

\begin{document}

%
\title{A Practical Survey on Emerging Threats from AI-driven Voice Attacks: How Vulnerable are Commercial Voice Control Systems?}

%
\author{Yuanda Wang,
Qiben Yan,
Nikolay Ivanov,
and Xun Chen

}


  
\maketitle

\begin{abstract}
The emergence of Artificial Intelligence (AI)-driven audio attacks has revealed new security vulnerabilities in voice control systems. While researchers have introduced a multitude of attack strategies targeting voice control systems (VCS), the continual advancements of VCS have diminished the impact of many such attacks. Recognizing this dynamic landscape, our study endeavors to comprehensively assess the resilience of commercial voice control systems against a spectrum of malicious audio attacks. Through extensive experimentation, we evaluate six prominent attack techniques across a collection of voice control interfaces and devices. Contrary to prevailing narratives, our results suggest that commercial voice control systems exhibit enhanced resistance to existing threats. Particularly, our research highlights the ineffectiveness of white-box attacks in black-box scenarios. Furthermore, the adversaries encounter substantial obstacles in obtaining precise gradient estimations during query-based interactions with commercial systems, such as Apple Siri and Samsung Bixby. Meanwhile, we find that current defense strategies are not completely immune to advanced attacks. Our findings contribute valuable insights for enhancing defense mechanisms in VCS. Through this survey, we aim to raise awareness within the academic community about the security concerns of VCS and advocate for continued research in this crucial area.
\end{abstract}



 
%
\IEEEpeerreviewmaketitle

\section{Introduction}
Voice control systems (VCS) have transformed the way users interact with computers by enabling interactions through natural language and voice commands. Conventionally, users can transmit speech audio to online automatic speech
recognition (ASR) service platforms, which in turn offer the corresponding text translations. Meanwhile, the development of deep learning has greatly advanced VCS technology, leading to more accurate speech recognition and natural language processing. As a result, the global VCS market is projected to reach a staggering 95.41 billion dollars by 2030~\cite{vuimarket}, indicating the immense potential and demand for this technology.
To optimize user experience and improve efficiency, VCS technology has been integrated into a wide variety of applications, such as smart devices, Internet of Things (IoT), and automobile systems. By allowing users to access information and control devices more conveniently, VCS offers an intuitive and seamless human-computer interaction solution. 

However, existing research has revealed that AI models, particularly those based on deep neural network (DNN) architectures, are susceptible to a range of  attacks. The vulnerability in DNNs has subsequently extended to AI-based VCS. 
To  investigate the robustness of VCS, it is imperative to have a thorough understanding the impacts of the diverse attacks and their respective defenses in real-world scenarios.
In this work, we identify four major research gaps that hinder the execution of successful practical attacks on commercial VCS. 

\noindent\textbf{Gap 1: Generic Black-box Attacks.}  
Most adversarial attacks adopt a white-box setup, which leverages gradient descent optimization to generate adversarial examples (AEs). This implies that adversaries require complete knowledge of the models within the VCS. However, this assumption may not hold in real-world scenarios.
Commercial VCS platforms, such as Apple Siri~\cite{siri_link}, Google Assistant~\cite{google_link}, Amazon Alexa~\cite{alexa_link}, and Samsung Bixby~\cite{bixby_link}, are black-box systems, where the attackers could only access the final output for a given input.  

To address the challenges of black-box attacks, recent studies have suggested that adversaries might optimize their AEs by querying the VCS to obtain an estimated gradient. However, this query-based gradient estimation approach has significant overheads. First, accurate gradient estimation demands a large number of queries ($\geq1000$); thus, querying the VCS to craft effective AEs becomes costly. Second, to prevent abuse, API services tend to restrict excessive queries from a single user, adding another layer of complexity to the query-based gradient estimation process. Meanwhile, executing such black-box attacks still requires prior knowledge such as the device model and its API, which is sometimes unobtainable in real-world scenarios.


\noindent\textbf{Gap 2: Physical Adversarial Audio Transmission.}
The generation of AEs often overlooks the complexities of physical world signal transmission, rendering them frequently ineffective in real-world settings during airborne transmission. On one hand, these perturbations experience energy loss during transmission, which reduces the attack's success rate. On the other hand, AEs are susceptible to signal distortions in complex environments. For instance, obstacles such as walls, floors, and furniture can reflect the sound, which causes echo effects in the signals captured by microphones. As a result, AEs designed for digital transmission cannot always work well in real-world settings. Some recent studies show that using a band-pass filter can help improve the performance of adversarial audio in airborne transmission~\cite{yakura2018robust}. Also, using room impulse response (RIR) filters can reduce the distortion caused by echoes. However, the success of over-the-air attacks varies and is not always reliable in different environmental conditions.

Moreover, the scope of adversarial attack threats is rather limited by its transmission media and surrounding environments. 
In many cases, adversaries need to be in close proximity to the target device. 
To increase attack range, some attackers have embedded adversarial perturbations within harmless music tracks~\cite{yuan2018commandersong, chen2020devil}. By hiding these perturbations within music or other non-speech audios, attackers can make them less noticeable. However, a major challenge for such attacks is their reliance on the victim who plays these manipulated audios. This means attackers cannot predict exactly when their attack would take place. Furthermore, for the attack to be effective, 
 the sound source must be positioned close to the target VCS to initiate and manipulate it.

A recent approach~\cite{liu2022evil} transmits malicious audio samples via telephone lines. Unlike the method of hiding malicious audio within music, over-the-line attacks can directly target specific individuals. For instance, attackers can send the malicious audio during a phone call or an online meeting. 
If the malicious audio is played aloud near the target device, the device could be compromised. 
However, launching attacks over telephone lines presents its own set of challenges: (i) The telephone line might introduce unforeseen distortions caused by the channel state, potentially reducing the likelihood of a successful attack. 
(ii) Data loss or unexpected delays can further impact the effectiveness of the malicious audio.

\noindent\textbf{Gap 3: Perceptibility.}
A key attribute of AEs is their ability to discreetly embed perturbations in benign audio, ensuring that it sounds natural and clear. However, these embedded perturbations can sometimes create noticeable distortions. To craft malicious audios that preserve naturalness, adversaries have employed various constraints on the perturbations. Many attacks use a constraint parameter $\epsilon$ to limit the amplitude of the perturbation, as described by $\Vert x - x^{*} \Vert_{k} < \epsilon$, where $x^{*}$ represents the AE.

However, $\epsilon$ alone is inadequate to optimally balance between the performance of the adversarial audio and its imperceptibility.
To better optimize the perturbation, some adversarial attacks introduce the perturbation amplitude into the loss function~\cite{qin2019imperceptible}.
Other attacks use signal-to-noise ratio (SNR) to constrain the perturbation strength and decrease its perceptibility~\cite{li2020advpulse, deng2022fencesitter}.

For white-box adversarial attacks that rely on GD, smaller perturbations can be effectively concealed within the benign audio. To further improve the stealthiness, a recent approach~\cite{schonherr2018adversarial} suggests embedding the perturbation near the initial signal frequencies, leveraging psychoacoustic properties to evade human detection. However, black-box attacks typically require more powerful perturbations to ensure success. In such scenarios, psychoacoustic-based methods become less effective, especially during over-the-air attacks. Consequently, victims could potentially detect anomalies in the audio and pinpoint the ongoing attack.

\noindent\textbf{Gap 4: Robustness to Defense.}
In response to the threats posed by adversarial attacks on VCS, researchers have devised various defense mechanisms to identify and neutralize malicious audio signals. One widely adopted approach is liveness detection. Given that malicious audios, when targeting real-world VCS devices are played through loudspeakers, the distinct acoustic properties of loudspeakers compared to human vocal cords can be leveraged to differentiate between audio from loudspeakers and live human speech. Smart speakers utilize microphone arrays to detect replayed audio signals, while some IoT devices can recognize differences between sounds transmitted through bone conduction and natural human voices~\cite{zhang2020viblive}. However, while liveness detection can identify malicious audios, it struggles to restore the original benign voice command from the adversarial samples.

Moreover, considering that the disturbances in adversarial examples are often carefully crafted, they can be effectively defended by certain signal processing techniques, such as down-sampling~\cite{hussain2021waveguard}. The diffusion model has gained attention due to its impressive generative capabilities, which offers a promising avenue for extracting genuine audio from adversarial samples~\cite{wu2022defending}. Furthermore, the rise of deepfake voice attacks has prompted the development of audio signal analysis techniques to detect and counteract them~\cite{zhang2021one, huang2021defending, wang2023vsmask}. 
Backdoor attacks have been developed to compromise SV models~\cite{deng2022fencesitter, guo2023masterkey}.
From an attacker's perspective, crafting attacks that can bypass emerging defense mechanisms is crucial. 
Conversely, for defenders, it is imperative to constantly update defenses, anticipating the evolving tactics of adaptive attackers.

The aforementioned challenges underline the complexity of comprehending and countering AI-driven threats aimed at VCS. Moreover, the effectiveness of these attacks and defenses cannot be universally assured across diverse real-world settings, emphasizing the importance of evaluating performance beyond controlled laboratory environments.

\noindent\textbf{Related Work:} Recent studies have addressed threats in speech recognition systems~\cite{abdullah2021sok, hu2019adversarial, wang2019adversarial, 9006862, zhang2022adversarial} and speaker recognition systems~\cite{tan2022adversarial, lan2022adversarial, das2020attacker}.
Additionally, the security of voice-user interfaces~\cite{bispham2019speech} and IoT devices~\cite{davis2020vulnerability, edu2020smart} has been thoroughly reviewed. 
A recent survey paper provides an exhaustive overview of security threats and countermeasures related to voice assistants~\cite{yan2022survey}. However, there is a lack of empirical evaluations of attacks and defenses in real-world scenarios. Consequently, the effectiveness of countermeasures against modern threats aimed at updated VCS remains unclear. 
With the rapid advancement of new technologies,  the security landscape of VCS is constantly evolving. This paper offers an revamped investigation and systematization of VCS security, specifically addressing AI-driven attacks.

Our contributions are as follows:
\begin{itemize}
\item \textbf{Classification.} We provide a comprehensive categorization of AI-driven attacks towards VCS. For each attack type, we characterize the specific attacks and highlight their respective strengths and weaknesses. We also enumerate typical countermeasures associated with these attacks.
\item \textbf{Real-world Evaluation.} We evaluate the robustness of popular commercial VCS, including Apple Siri, Google Assistant, and Samsung Bixby. Contrary to prevailing assumptions, our measurements reveal that these systems are more resilient than anticipated, with over-the-air attacks proving particularly challenging against the latest VCS iterations.
\item \textbf{Recommendations.} Drawing from our survey and empirical results, we identify promising avenues for future research. Our insights aim to guide both attackers and defenders, potentially catalyzing further research to enhance VCS security.
\end{itemize}
\section{Background}
In this section, we briefly introduce VCS, SV, and ASR technologies.
\vspace{-10pt}
\subsection{Voice Control Systems}
In smart homes, VCS often serve as the central interface, enabling users to remotely operate devices, including lights, doors, and air conditioners. Numerous manufacturers have launched their VCS products, as listed in Table~\ref{Tab: products}. These systems remain dormant until triggered by specific wake words, such as ``Hey Siri" or ``Hello Google." Once activated, the subsequent voice inputs are relayed to their APIs, resulting in actionable tasks. Following this, the VCS execute the commands and offer feedback.
\vspace{-10pt}

\begin{table}
\centering
\caption{Popular voice assistants on mobile devices.}
\label{Tab: products}
\begin{tabular}{cccc} 
\toprule
\textbf{Manufacturer} & \textbf{Voice Assistant} & \textbf{API} & \begin{tabular}[c]{@{}c@{}}\textbf{Wake-up }\\\textbf{Word}\end{tabular}  \\ 
\midrule
Apple                 & Siri~                    & N/A          & \textit{(Hey) Siri}                                                       \\
Samsung               & Bixby~                   & N/A          & \textit{Hi Bixby}                                                         \\
Microsoft             & Cortana~                 & Azure        & \textit{Cortana}                                                          \\
Google                & Google Assistant~        & Google ASR   & \textit{Hey Google}                                                       \\
Amazon                & Alexa~                   & AWS          & \textit{Alexa}                                                            \\
\bottomrule
\end{tabular}
\end{table}

\begin{figure*}[t]
    \centering
    \includegraphics[width=0.9\textwidth]{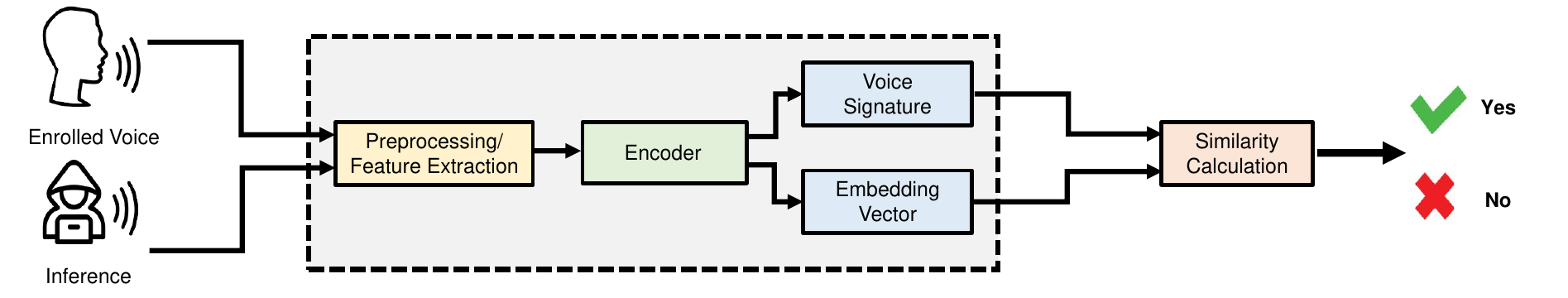}
    \caption{Speaker Verification (SV) framework.}
    \label{fig:SV}
\end{figure*}

\subsection{Speaker Verification}
Speaker verification (SV) is essential to ensure the security and user privacy of VCS. It verifies the speaker's identity before processing voice commands. A typical SV system has five modules: feature extraction, background model, speaker model, scoring, and decision~\cite{chen2021real}, as shown in Fig.~\ref{fig:SV}.
The feature extraction module transforms a raw speech signal into acoustic features that capture the signal's characteristics. 
Different acoustic features have been proposed, for example, Mel-Frequency Cepstral Coefficients (MFCC), Spectral Subband Centroid (SSC), and Perceptual Linear Predictive (PLP). 

For SV models, i-vector based Probabilistic Linear Discriminant Analysis (PLDA) is a mainstream speaker verification approach to evaluate the similarity of enrolled voice and the input voice~\cite{snyder2019speaker, dehak2010front}.
Meanwhile, SV models based on Gaussian Mixture Model (GMM) can achieve similar accuracy as i-vector PLDA models, especially for short input speech~\cite{reynolds2000speaker}.
Nowadays, DNN based models are widely deployed, for example,  Time Nelay neural Network (TDNN) method has demonstrated the best SV accuracy~\cite{speechbrain, dawalatabad2021ecapa}.

In SV systems, the primary distinction from speaker recognition (SR) is that SV does not seek to classify or distinguish among various speakers. 
SV's main objective is to authenticate a specific speaker based on a predefined threshold.
In VCS, the SV process is generally text-dependent to avoid unintended activation. 
\vspace{-10pt}

\subsection{Automatic Speech Recognition}
Upon verifying the speaker's identity, users can issue voice commands to remotely operate the VCS. These commands are processed by ASR systems. Much like SV, a standard ASR system undergoes a three-step procedure that includes pre-processing, feature extraction, and language modeling. However, the key difference is that ASR focuses on extracting content from speech, rather than features related to the speaker's identity.

\begin{figure*}[t]
    \centering
    \includegraphics[width=0.9\textwidth]{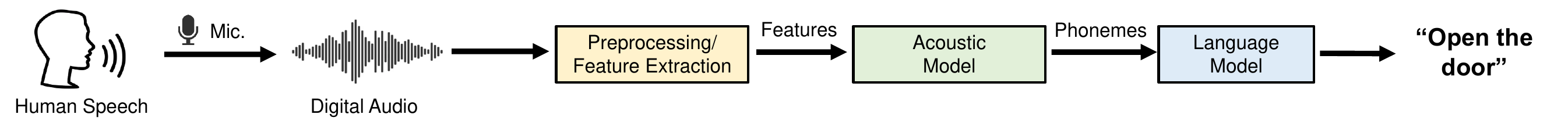}
    \caption{Automatic Speech Recognition (ASR) framework.}
    \label{fig:ASR framework}
\end{figure*}

 A typical ASR model framework is shown in Fig.~\ref{fig:ASR framework}. 
 Hidden Markov Model (HMM) was widely used for early ASR systems~\cite{gales1998maximum}. Recently, with advancements in machine learning, DNNs, particularly recurrent neural networks (RNNs), have become increasingly prominent  in speech recognition, due to their ability to capture contextual information~\cite{graves2013speech}.
After signal transformations, ASR can extract speech features and input them to the acoustic model.
The audio signals will be transcribed to utterances and then phonemes. 
In the next step, the language model will predict the text content in the human speech.
Finally, the Spoken Language Understanding (SLU) model extracts the key information in the speech and initializes corresponding operations indicated in the voice command.
\vspace{-10pt}

\begin{figure*}[t]
    \centering
    \includegraphics[width=16cm]{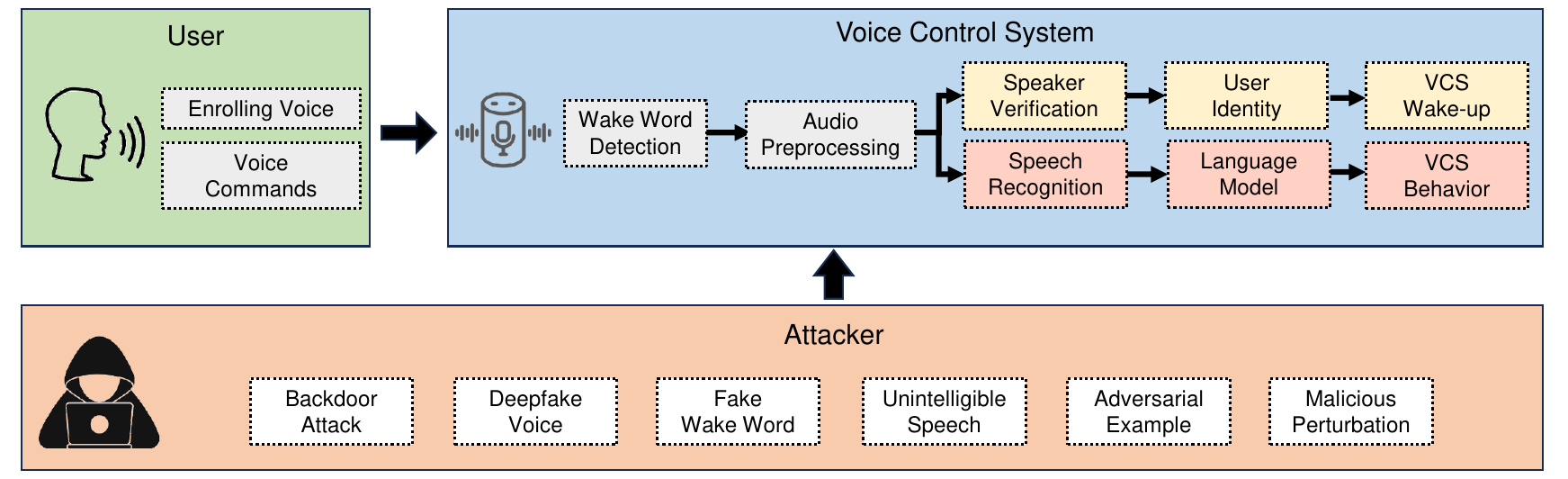}
    \caption{A framework of VCS workflow and potential vulnerabilities.}
    \label{fig:threat_model}
\end{figure*}

\begin{table*}
\centering
\normalsize
\caption{Attack methods towards SV models.}
\label{Tab: attack_for_speaker}
\begin{tabular}{|c|c|c|c|c|c|c|} 
\hline
Type                                 & Attack       & Method         & Knowledge & Target Model & Airborne & Carrier              \\ 
\hline
\multirow{5}{*}{Deepfake Voice}      & AdaIN-VC~\cite{chou2019one}    & VC             & Black     & N/A            & Yes      & Synthesized voice    \\ 
\cline{2-7}
                                     & AutoVC~\cite{qian2019autovc}      & VC             & Black     & N/A            & Yes      & Synthesized voice    \\ 
\cline{2-7}
                                     & SV2TTS~\cite{jia2018transfer}      & TTS            & Black     & N/A            & Yes      & Synthesized voice    \\ 
\cline{2-7}
                                     & plat.ht~\cite{playht}      & TTS            & Black     & N/A            & Yes      & Synthesized voice    \\ 
\cline{2-7}
                                     & resemble.ai~\cite{resembleai}  & TTS            & Black     & N/A            & Yes      & Synthesized voice    \\ 
\hline
\multirow{5}{*}{Adversarial Example} & FakeBob~\cite{chen2021real}     & Transferability & Black     & Azure        & Yes  & Adversarial example     \\ 
\cline{2-7}
                                     & SirenAttack~\cite{du2020sirenattack} & PSO            & Black     & 6 models     & No  & Adversarial example      \\ 
\cline{2-7}
                                     & AdvPulse~\cite{li2020advpulse}    & GD             & White     & x-vector     & Yes  & Perturbation     \\ 
\cline{2-7}
                                     & End-to-end~\cite{wang2019adversarial}  & FGSM           & White     & DNN E2E      & No   & Adversarial example     \\ 
\cline{2-7}
                                     & VMask~\cite{zhang2020voiceprint}       & SGD            & Black     & Azure/Siri   & Yes   & Adversarial example    \\
\hline
\end{tabular}
\end{table*}

\begin{table*}
\centering
\caption{Existing attacks towards ASR models. We classify the attacks according to their feasibility under digital (D), airborne (A) and cellular (C) transmission, as well as the attack signals' intelligibility (INT) and their susceptibility to noise (SUS).}
\label{Tab: attacks_for_ASR}
\begin{tabular}{|c|c|c|c|c|c|c|c|c|c|c|} 
\hline
\multirow{2}{*}{Type}                                                           & \multirow{2}{*}{Attack} & \multirow{2}{*}{Method} & \multirow{2}{*}{Gradient-free}& \multirow{2}{*}{Target Model} & \multirow{2}{*}{Targeted} & \multicolumn{3}{c|}{Transmission} & \multirow{2}{*}{INT} & \multirow{2}{*}{SUS}  \\ 
\cline{7-9}
                                                                                &                         &                         &                            &                               &                           & D & A & C               &         &        \\ 
\hline
\multirow{3}{*}{\begin{tabular}[c]{@{}c@{}}Unintelligible \\Audio\end{tabular}} & Cocaine noodles~\cite{vaidya2015cocaine}         & MFCC-Rec                & \CIRCLE                      & Google Now                    & \CIRCLE                  &  \CIRCLE     &\CIRCLE      &        \Circle             &     \Circle    &  \CIRCLE      \\ 
\cline{2-11}
& Hidden Commands   ~\cite{carlini2016hidden}      & MFCC-Rec                & \CIRCLE                      & Google Now                    & \CIRCLE                  &  \CIRCLE    &  \CIRCLE    &      \Circle               &      \Circle   &  \CIRCLE      \\ 
\cline{2-11}
& Practical ~\cite{abdullah2019practical}              & Signal Pro.             & \CIRCLE                      & 12 Models                     & \CIRCLE                  &     \CIRCLE    &  \CIRCLE    &      \Circle                     &       \Circle   &  \CIRCLE        \\ 
\hline
\multirow{3}{*}{\begin{tabular}[c]{@{}c@{}}Malicious \\Perturbation\end{tabular}}                                                                               & AdvPulse     ~\cite{li2020advpulse}           & GD                      & \Circle                      & KWS                           & \CIRCLE                  &       \CIRCLE &  \CIRCLE    &       \Circle              &  \Circle       &  \CIRCLE        \\ 
\cline{2-11}
 & VoiceCamo~\cite{chiquier2021real}               & Predictive model        & \Circle                      & DeepSpeech                    & \Circle                &       \CIRCLE &  \CIRCLE    &       \Circle              &  \Circle       &  \CIRCLE        \\ 
\cline{2-11}
& SpecPatch~\cite{guo2022specpatch}           & GD                      & \Circle                      & DeepSpeech2                   & \CIRCLE             &      \CIRCLE &  \CIRCLE    &       \Circle              &  \Circle       &  \CIRCLE        \\ 
                                                                               
\hline
\multirow{24}{*}{\begin{tabular}[c]{@{}c@{}}Adversarial\\Examples\end{tabular}} & Houdini~\cite{cisse2017houdini}                 & GD                      & \Circle \ \CIRCLE                & DeepSpeech                    & \Circle                &      \CIRCLE    &  \CIRCLE    &      \Circle                     &     \CIRCLE    &   \Circle     \\ 
\cline{2-11}
                                                                                & Alzantot et al. ~\cite{alzantot2018did}        & GA                      & \CIRCLE                      & KWS                           & \CIRCLE                  &     \CIRCLE &  \Circle    &       \Circle              &  \CIRCLE       &  \Circle      \\ 
\cline{2-11}
                                                                                & Iter et al. ~\cite{iter2017generating}           & GD                      & \Circle                      & WaveNet                       & \CIRCLE                  &     \CIRCLE &  \Circle    &       \Circle              &  \CIRCLE       &  \Circle           \\ 
\cline{2-11}
                                                                                & Carlini et al.  ~\cite{carlini2018audio}       & GD                      & \Circle                      & DeepSpeech                    & \CIRCLE                  &       \CIRCLE &  \Circle    &       \Circle              &  \CIRCLE       &  \Circle        \\ 
\cline{2-11}
                                                                                & Schonherr et al.~\cite{schonherr2018adversarial}        & GD                      & \Circle                      & Kaldi                         & \CIRCLE                  &      \CIRCLE &  \Circle    &       \Circle              &  \CIRCLE       &  \Circle        \\ 
\cline{2-11}
                                                                                & Yakura et al.~\cite{yakura2018robust}           & GD                      & \Circle                      & DeepSpeech                    & \CIRCLE                  &       \CIRCLE &  \CIRCLE    &       \Circle              &  \Circle       &  \Circle        \\ 
\cline{2-11}
                                                                                & Qin et al.~\cite{qin2019imperceptible}              & GD                      & \Circle                      & Lingvo                        & \CIRCLE                  &       \CIRCLE &  \CIRCLE    &       \Circle              &  \CIRCLE       &  \Circle        \\ 
\cline{2-11}
                                                                                & Taori et al.   ~\cite{taori2019targeted}         & GD                      & \RIGHTcircle                       & DeepSpeech                    & \CIRCLE                  &       \CIRCLE &  \Circle    &       \Circle              &  \CIRCLE       &  \Circle        \\ 
\cline{2-11}
                                                                                & SirenAttack ~\cite{du2020sirenattack}            & GD/PSO                  & \Circle \ \CIRCLE                & DeepSpeech                    & \CIRCLE                  &       \CIRCLE &  \Circle    &       \Circle              &  \CIRCLE       &  \Circle        \\ 
\cline{2-11}
                                                                                & Neekhara et al. ~\cite{neekhara2019universal}        & GD                      & \Circle                      & WaveNet                       & \Circle                &       \CIRCLE &  \Circle    &       \Circle              &  \CIRCLE       &  \Circle        \\ 
\cline{2-11}
                                                                                & Kwon et al.    ~\cite{kwon2019selective}         & GD                      & \Circle                      & DeepSpeech                    & \CIRCLE                  &       \CIRCLE &  \Circle    &       \Circle              &  \CIRCLE       &  \Circle        \\ 
\cline{2-11}
                                                                                & Li et al.~\cite{li2019adversarial}               & GD                      & \RIGHTcircle                       & Alexa                         & \Circle                &       \CIRCLE &  \Circle    &       \Circle              &  \Circle       &  \Circle        \\ 
\cline{2-11}
  & Liu et al. ~\cite{liu2020weighted}             & GD                      & \Circle                      & DeepSpeech                    & \CIRCLE                  &       \CIRCLE &  \Circle    &       \Circle              &  \CIRCLE       &  \Circle        \\ 
\cline{2-11}
   & Imperio ~\cite{schonherr2020imperio}                & GD                      & \Circle                      & Kaldi                         & \CIRCLE                  &       \CIRCLE &  \CIRCLE    &       \Circle              &  \CIRCLE       &  \Circle        \\ 
\cline{2-11}
 & Szurley et al.  ~\cite{szurley2019perceptual}        & GD                      & \Circle                      & DeepSpeech                    & \CIRCLE                  &       \CIRCLE &  \CIRCLE    &       \Circle              &  \CIRCLE       &  \Circle        \\ 
\cline{2-11}
 & Metamorph ~\cite{chen2020metamorph}              & GD                      & \Circle                      & DeepSpeech                    & \CIRCLE                  &      \CIRCLE &  \CIRCLE    &       \Circle              &  \CIRCLE       &  \Circle        \\ 
\cline{2-11}
& Chang et al. ~\cite{chang2020audio}           & RNN                     & \Circle                      & KWS                           & \CIRCLE                  &       \CIRCLE &  \CIRCLE    &       \Circle              &  \CIRCLE       &  \Circle        \\ 
\cline{2-11}
 & Wang et al.~\cite{wang2020targeted}             & GAN                     & \Circle                      & WideResNet                    & \CIRCLE                  &       \CIRCLE &  \Circle    &       \Circle              &  \CIRCLE       &  \Circle        \\ 
\cline{2-11}
 & CommandSong~\cite{yuan2018commandersong}             & Transferability          & \Circle \ \CIRCLE                & Kaldi                         & \CIRCLE                  &       \CIRCLE &  \CIRCLE    &       \Circle              &  \Circle       &  \CIRCLE        \\ 
\cline{2-11}
& Kenansville~\cite{abdullah2021hear}          & Removing Frequency          & \CIRCLE                      & 7 models                      & \Circle                &      \CIRCLE &  \CIRCLE    &       \CIRCLE              &  \CIRCLE       &  \Circle        \\ 
\cline{2-11}
 & Devil's whisper~\cite{chen2020devil}       & Transferability                      & \CIRCLE                      & 4 commercial                  & \CIRCLE                  &      \CIRCLE &  \CIRCLE    &       \Circle              &  \Circle       &  \CIRCLE        \\ 
\cline{2-11}
 & OCCAM~\cite{zheng2021black}                 & Transferability           & \CIRCLE                      & 4 commercial                  & \CIRCLE                  &      \CIRCLE &  \CIRCLE    &       \Circle              &  \Circle       &  \CIRCLE        \\ 
\cline{2-11}
 & TAINT~\cite{liu2022evil}                 &   GD                     & \CIRCLE                      & 4 Models                      & \CIRCLE                          &     \CIRCLE &  \CIRCLE    &       \CIRCLE              &  \CIRCLE       &  \Circle        \\ 
\cline{2-11}
 & SMACK ~\cite{yusmack}                  & Prosody Manipulation                         & \CIRCLE                      & 5 Models                      &  \CIRCLE                         &     \CIRCLE &  \CIRCLE    &       \Circle              &  \CIRCLE       &  \Circle        \\
\hline
\end{tabular}
\begin{tablenotes}\footnotesize
\centering
\item Gradient-free characteristic, \Circle: white-box attack, \CIRCLE:~black-box attack, \RIGHTcircle:~grey-box attack.
\end{tablenotes}
\end{table*}
\section{AI-driven Attacks}
With the progression of AI technologies, VCS have become increasingly capable in handling speaker identity verification and speech recognition. 
In this section, we present a comprehensive framework to illustrate potential AI-driven attacks targeting VCS, which is shown in Fig.~\ref{fig:threat_model}. 
The attacks targeting SV and ASR models are respectively listed in Table~\ref{Tab: attack_for_speaker} and Table~\ref{Tab: attacks_for_ASR}.


\subsection{Deepfake Voice Attacks}\label{deepfakeattack}


VCS are crucial in the IoT ecosystem, making them enticing targets for malicious actors. 
To protect VCS against malicious commands from adversaries, SV models have been deployed to verify the user's identity.
When VCS are inactive, they do not recognize any voice commands. 
Therefore, the attackers have to bypass the SV models~\cite{wang2022ghosttalk}.

\begin{figure}[htbp]
    \centering
    \subfigure[Voice conversion synthesis.]{\includegraphics[width=0.47\textwidth]{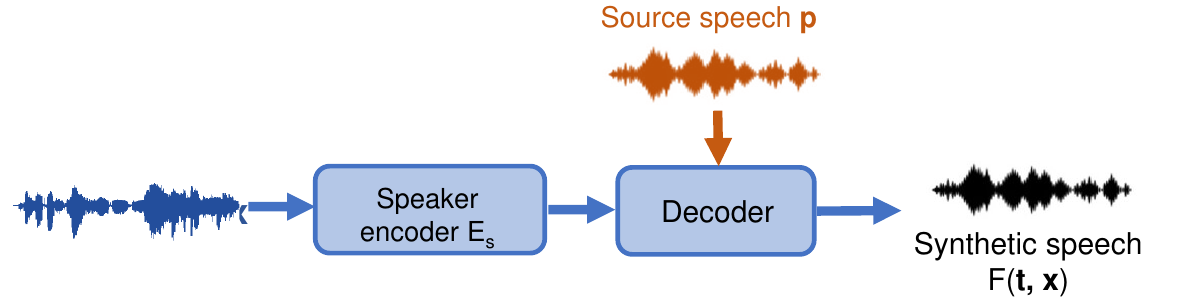}}\label{fig:vc}
    \subfigure[Text-to-speech synthesis.]{\includegraphics[width=0.47\textwidth]{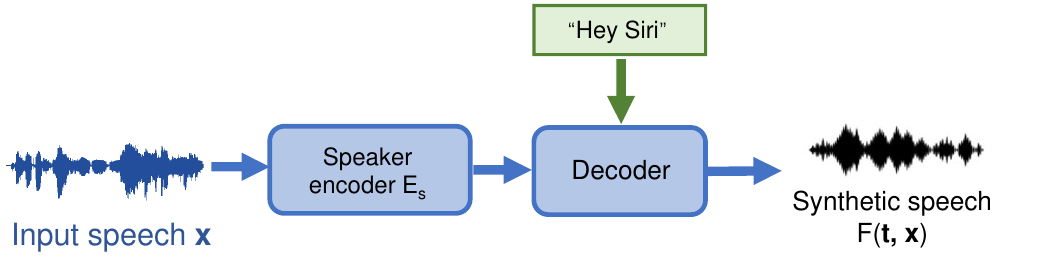}}\label{fig:tts}
    \caption{Voice synthesis frameworks.}
    \vspace{-10pt}
    \label{fig:voice_synthesis}
\end{figure}

Deepfake voice attacks aim to produce artificial speech that mimics real human voices, with the intent to deceive SV models. There are two primary methods for generating deepfake voices: text-to-speech (TTS) and voice conversion (VC).
TTS models need speech samples from the targeted speaker. 
These models capture voice features from these samples and incorporate them into the synthesized speech, while the content of the speech is derived from a provided text.
On the other hand, VC models also capture voice patterns from the target speaker's samples, but they derive the linguistic content from a source speech, rather than text.
For both TTS and VC approaches, the result is a synthetic speech in the target speaker's voice. 
Fig.~\ref{fig:voice_synthesis} illustrates a general voice synthesis framework that encompasses both TTS and VC methods.
For TTS and VC models, the speech content embeddings are respectively extracted by the content encoders $E_{tts}$ or $E_{c}$ from text $\textbf{t}$ or speech $\textbf{p}$.
The speaker encoder $E_{s}$ encodes the victim speaker characteristics $E_{s}(\textbf{x})$, where $\textbf{x}$ is a speech sample from the victim speaker $x$.
Finally, the decoder will generate a synthetic speech $F(\textbf{x}, \textbf{p})$ or $F(\textbf{x}, \textbf{t})$ with voice features from an arbitrary victim speaker $x$.
Deepfake attack requires no prior-knowledge about the SV models, making it a generic attack against all VCS.

\subsection{Backdoor Attacks}\label{backdoor}
Recent studies highlight the vulnerability of SV models to backdoor attacks~\cite{zhai2021backdoor, shi2022audio,  guo2023masterkey}.
In this type of attack, adversaries embed specific backdoor triggers into the training dataset, which can subsequently manipulate the model's behavior.
During the model's inference phase, when these triggers are present in an input, they can cause the model to make incorrect predictions or classifications.
However, it is worth noting that for commercial VCS platforms, attackers typically do not have direct access to the training datasets.
This lack of access significantly complicates the practical execution of such backdoor attacks in real-world scenarios.

FenceSitter~\cite{deng2022fencesitter} is a backdoor attack that incorporates an imperceptible backdoor trigger during the enrollment phase of SV. Unlike conventional backdoor attacks, this approach only embeds the trigger in the enrolled speech commands. 
The crafted trigger is both synchronization-free and compatible with black-box settings, thereby enhancing its applicability in real-world scenarios.

\subsection{Fake Wake-up Word Attacks}
In smart devices, VCS are usually activated by specific wake-up words such as ``Hey Siri" or ``Hello Google." Yet, this triggering mechanism is not entirely secure. Due to the phonetic similarities between certain sounds and actual wake-up words, VCS can be susceptible to fake wake-up word attacks. By exploiting this, attackers can induce unintended activation in the real world. 
A recent study, FakeWake~\cite{chen2021fakewake}, investigates the potential risks of fake wake words in VCS and assesses their implications for user privacy. 
Using a fuzzing approach, FakeWake identifies multiple false wake words that could cause inadvertent activation. 
The deceptive nature of fake wake words means they can covertly activate VCS, giving adversaries the opportunity to issue malicious commands without the user's knowledge.

\subsection{Unintelligible Speech Attacks}
\begin{figure*}[t]
    \centering
    \subfigure[Raw audio signal.]{\includegraphics[height=2.3cm]{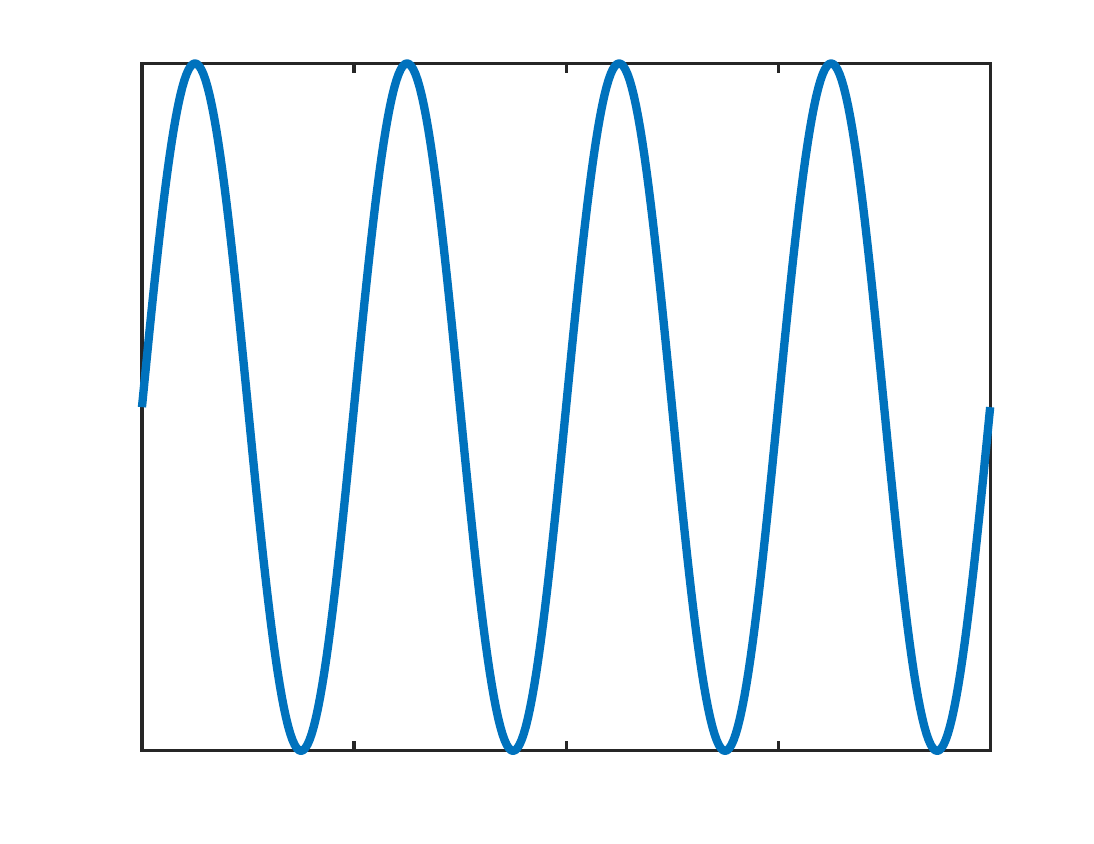}\label{fig:y}}\quad
    \subfigure[Phase shift.]{\includegraphics[height=2.3cm]{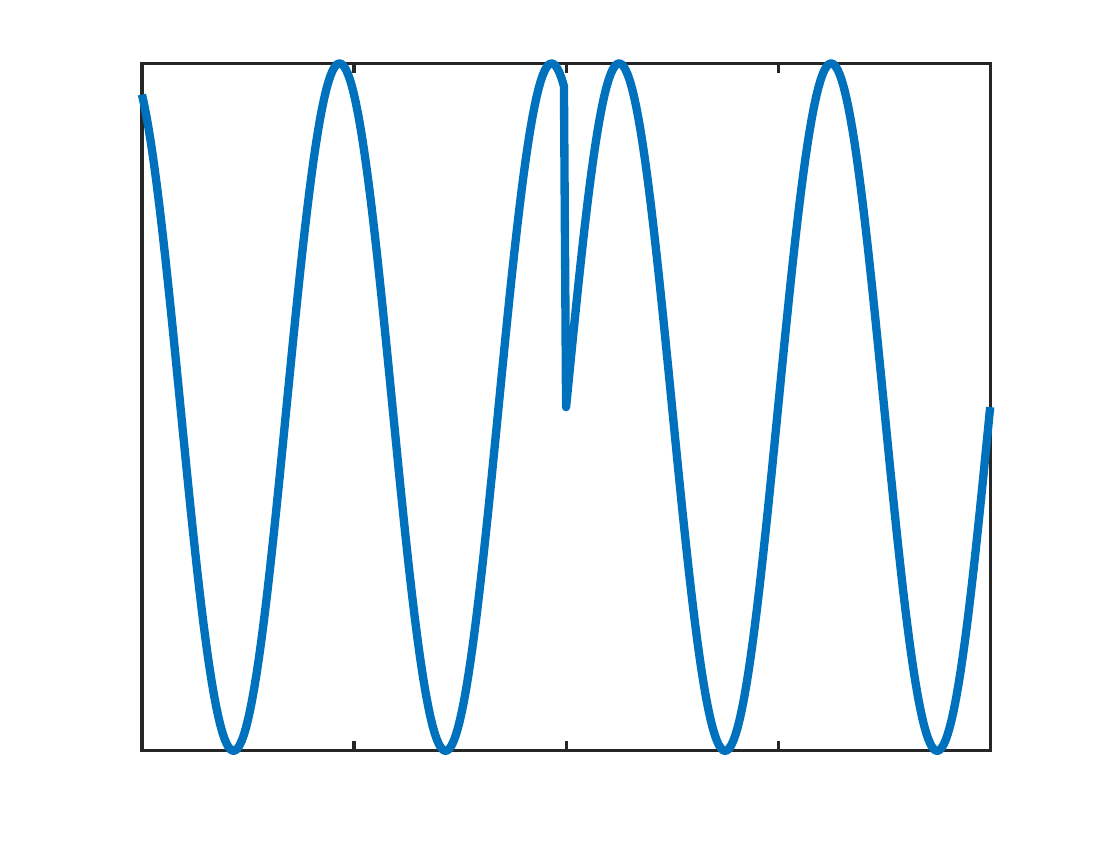}\label{fig:y2}}\quad
    \subfigure[Signal inversion.]{\includegraphics[height=2.3cm]{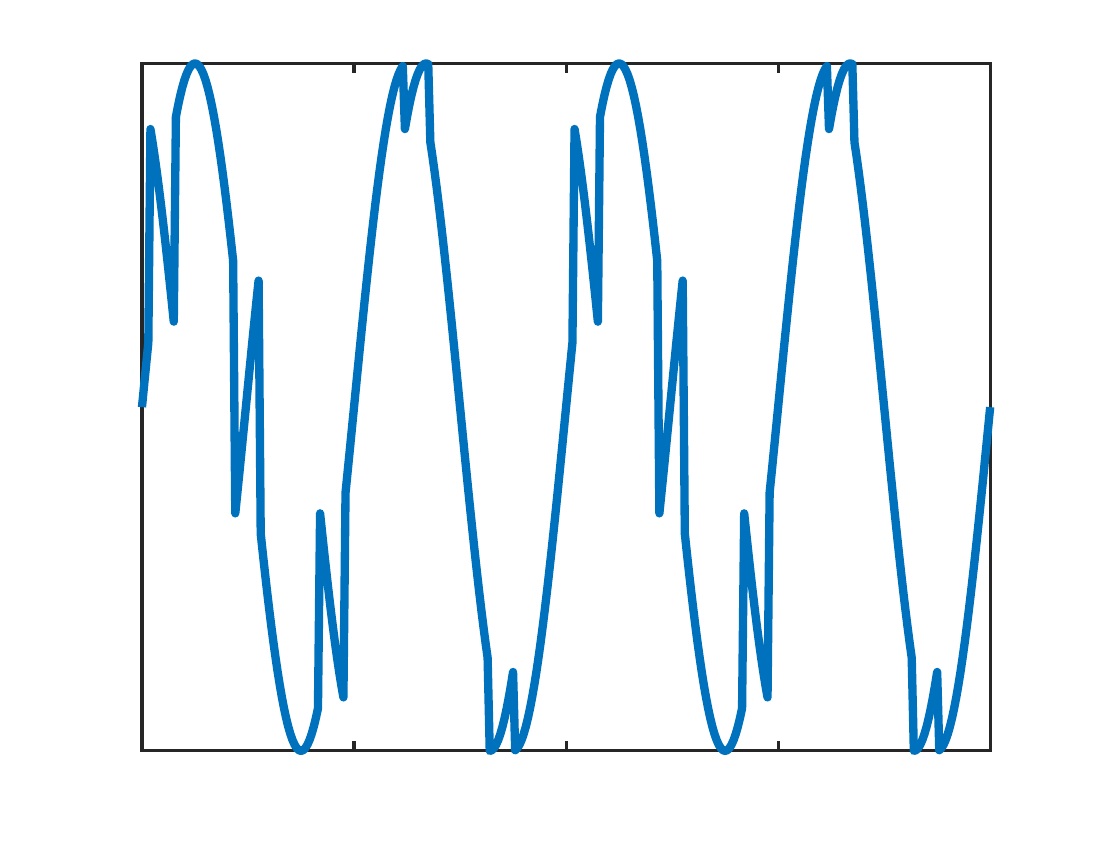}\label{fig:y1}}
    \subfigure[Time scaling.]{\includegraphics[height=2.3cm]{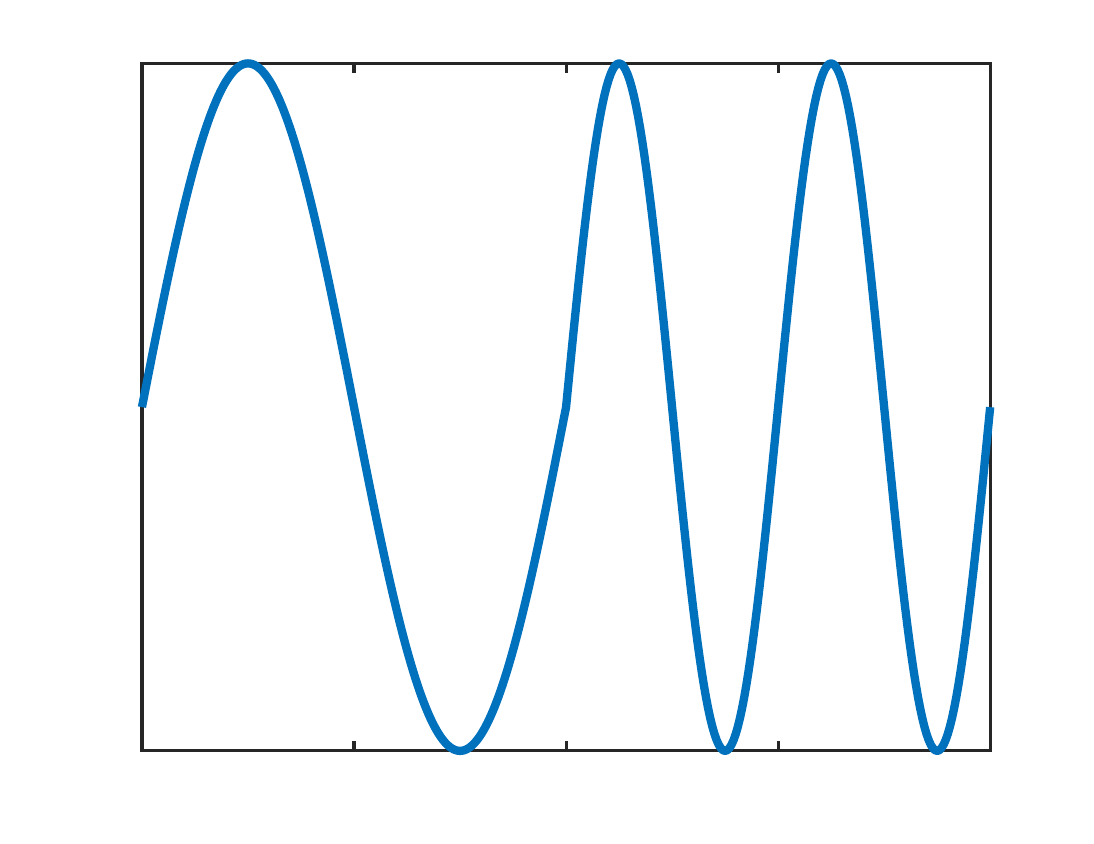}\label{fig:y3}}
    \subfigure[High-frequency noise.]{\includegraphics[height=2.3cm]{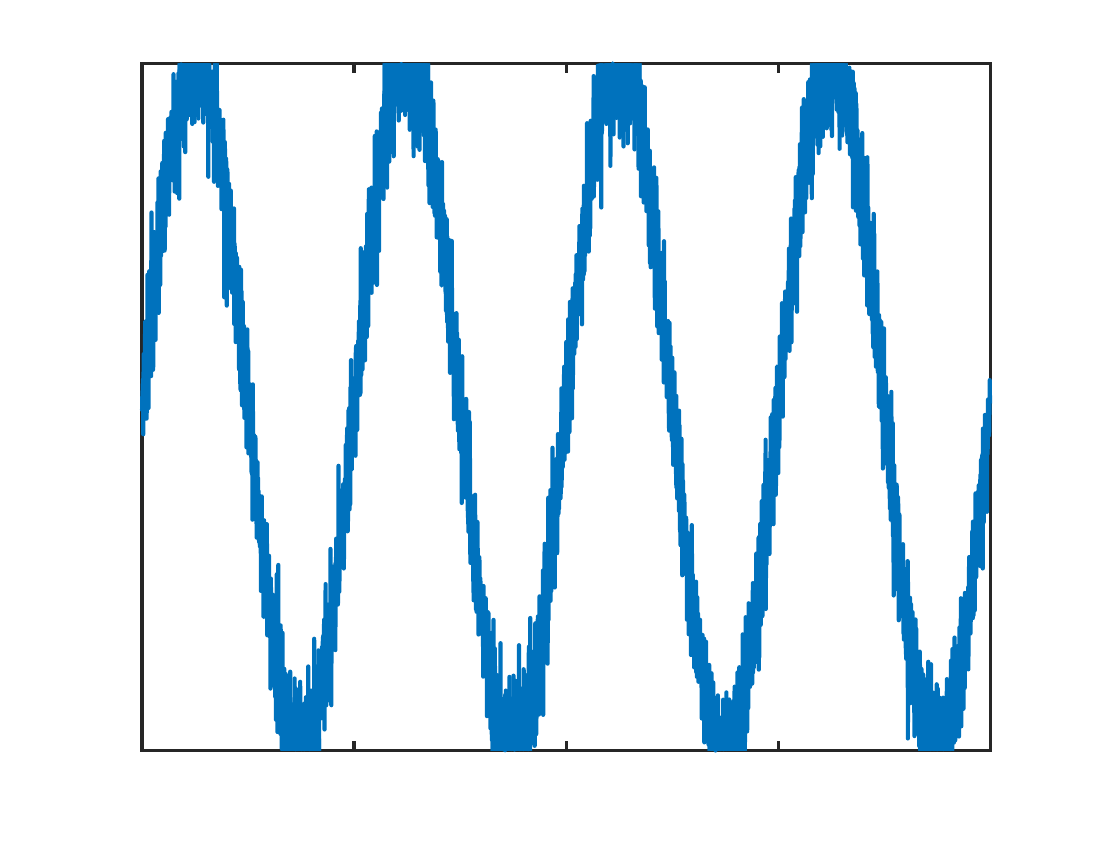}\label{fig:y4}}
    \caption{Signal transformation approaches for generating unintelligible speech.}
    \label{fig: unintelli}
\end{figure*}

Attackers can exploit VCS by introducing noise that, while inaudible or nonsensical to human listeners, can be interpreted as clear commands by ASR models. 
This type of attack primarily aims at the signal processing component of ASR systems. 
A hidden voice command attack, proposed by Calini et al.~\cite{carlini2016hidden}, is one example of unintelligible speech attacks. It involves launching incomprehensible speech by reconstructing MFCC features. Additionally, as shown in Fig.~\ref{fig: unintelli}, various signal transformation techniques, including phase shifting, signal inversion, time scaling, or adding high-frequency noise, can produce such unintelligible speech that can still be interpreted by ASR models.

\subsection{Adversarial Examples}
Carlini and Wagner first reveal that neural networks are vulnerable to adversarial image examples~\cite{7958570}.
Similarly, attackers can embed extremely subtle perturbations in audio samples, causing ASR or SV  models to produce incorrect predictions, while these adversarial examples remain undetectable to human ears.
We list typical methods to optimize AEs.

\noindent\textbf{Gradient Descent (GD).}
Early adversarial attacks usually require complete access to the target DNN model, which is known as white-box attack.
Suppose the target SV or ASR model is $f$, the adversarial perturbation signal is $\delta$, the optimization of AEs can be illustrated in Problem (\ref{adv_opt}).
\begin{equation}
\mathop{\textnormal{minimize}}\limits_{\delta} \quad \mathcal{L}(f(x+\delta), f(y)), \quad \textbf{s.t.} \quad \Vert \delta \Vert_{k} < \epsilon.
\label{adv_opt}
\end{equation}
where the perturbation $\delta$ amplitude is constrained by its $k$-norm.
$\mathcal{L}$ is the loss function, such as Mean Square Error (MSE), Cross Entropy or Kullback–Leibler (KL) divergence.
When targeting speech recognition systems, the adversaries usually use KL divergence loss, and MSE loss for speaker recognition models.
The adversaries can optimize $\delta$ according to the gradient $\nabla_{\delta}\mathcal{L}$.
A commonly applied method is Fast Gradient Sign Method (FGSM) which quickly optimizes AEs as follows:
\begin{equation}
 x^{*} = x + \varepsilon \textnormal{sgn} (\nabla_{x} L(\theta, x, y)), 
\end{equation}
where $x$ is the original audio signal and $y$ is the target label.
Compared with FGSM, Projected Gradient Descent (PGD) demonstrates greater efficacy, which can be described by the following optimization function:
\begin{equation}
    x^{t+1} = x^{t} + \alpha \cdot \textnormal{sgn} (\nabla_{x} L(\theta, x, y)),
\end{equation}
where $\alpha$ denotes the step length of the optimization.

\noindent\textbf{Gradient Estimation (GE).}
In reality, obtaining full access to SV or ASR models within commercial VCS is often impractical. In most cases, users can only query the model and obtain the final prediction results. To address this issue, some adversaries attempt to estimate the gradient by querying the model, which then facilitates the update of perturbation $\delta$:
\begin{equation}
\hat{\nabla f(x)} = \frac{f(x + \delta) - f(x)}{\delta}.
\end{equation}
Moreover, particle swarm optimization (PSO) and genetic algorithm (GA) are applied to launch gradient-free black-box attacks and achieve superior performance when compromising black-box models~\cite{du2020sirenattack}.
Nevertheless, for black-box attacks, the main challenge is that querying commercial VCS can be costly.
For example, OCCAM~\cite{zheng2021black} requires up to 30,000 queries to successfully compromise a commercial API.
Meanwhile, these black-box attacks typically rely on more than just the model's output; they often need additional predictions, such as probability distributions or similarity scores, to approximate the gradient. 
Consequently, accurately estimating the gradient from pure black-box models without incurring excessive query costs remains a significant challenge.

\noindent\textbf{Transferability.}
On the other hand, attackers could attempt to address the gradient-free challenge using transferable models.
First, they prepare a training set and obtain labels for the data by querying the target black-box model.
Next, they train a smaller substitute model based on these labels.
Once trained, this substitute model can be used to generate effective AEs.

However, this method faces several challenges. On one hand, commercial VCS often employ large-scale models to achieve optimal performance. For attackers, replicating such a model's scale and complexity is challenging, resulting in an inaccurate substitute model. On the other hand, there is also a risk of the substitute model overfitting to the training data. This means that while it might perform well on its training data, its effectiveness can drop significantly on new, unseen data.
CommanderSong~\cite{yuan2018commandersong} utilizes attack transferability to produce AEs. Instead of merely deriving the gradient from a substitute model, CommanderSong introduces a probability density function ID (\emph{pdf-id}) sequence match. This method aims to minimize the discrepancy between the AE and target benign speech. Additionally, CommanderSong conceals the perceptibility of the AEs by embedding them into music.
Devil's whisper~\cite{chen2020devil}  employs a large-scale white-box model to craft AEs using coarse-grained features. It then refines these AEs using a smaller substitute model to enhance their transferability.
OCCAM~\cite{zheng2021black} further refines AE generation by incorporating the AdaBelief~\cite{zhuang2020adabelief} method. This approach boosts the transferability of AEs by gradually decreasing the learning rate. Remarkably, compared to previous methods, OCCAM manages to execute an adversarial attack without any interaction with the target model.

\noindent\textbf{Prosody Manipulation.}
Existing attacks aim to manipulate the ASR by embedding adversarial perturbations into benign speech samples. However, finding the optimal balance between attack perceptibility and effectiveness remains a significant challenge. 
SMACK~\cite{yusmack} proposes a promising solution to this dilemma, as it introduces an adversarial attack that adjusts the prosody of human speech. Since this method refrains from adding extra noise and only modifies certain phonetic details, it can compromise ASR systems without heightening human perceptibility. 
However, its effectiveness is constrained as it only induces false transcriptions that are similar to the true ones.
\subsection{Malicious Perturbation}
AEs require specific carriers to conceal malicious commands, which presents a challenge in real-world attack scenarios where the perturbation signal strength is constrained. Additionally, AEs are susceptible to degradation from  user interference, environmental noise and propagation loss. To mitigate the limitations of AEs, recent advancements suggest that attackers could integrate malicious perturbations into VCS in tandem with users' benign commands. Malicious perturbations have been shown to be more resilient in airborne scenarios 
compared with AEs.

AdvPulse~\cite{li2020advpulse} utilizes a concise ``pulse" perturbation signal to force the ASR model in generating the target command regardless of the benign input. 
The model achieves synchronization-free perturbation which is formulated through the following optimization:
\begin{equation}
\begin{split}
      &\textnormal{minimize} \quad  \mathcal{L}(f(x + \delta_{i}), y_{t}) + \alpha \cdot \Vert \delta \Vert_{2},\\  
      &\textnormal{subject to} \quad  \Vert \delta \Vert_{2} \leq \epsilon,
\end{split}
\end{equation}
where $\delta_{i}$ is the same pulse signal with different random time shifts.
As the optimization is based on a large amount of speech carriers and time delays, AdvPulse achieves a robust, transferable attack that maintains its effectiveness across various benign inputs. 
However, these attacks face limitations when attempting to compromise ASR models in generating longer commands consisting of multiple words. 
Additionally, during an attack, if the user keeps speaking, the ASR system can still recognize the ending of benign commands, thereby disabling the malicious commands generated by AdvPulse.
In contrast, SpecPatch~\cite{guo2022specpatch} specifically targets RNN-based ASR models. Compared to AdvPulse, SpecPatch can more effectively manipulate the RNN model using a short patch as a malicious perturbation. This patch not only alters the affected audio output but also deceives the RNN model, leading to misinterpretation of subsequent audio or even muting upcoming speech. 

In addition to GD-based attacks, VoiceCamo~\cite{chiquier2021real} introduces a real-time method for generating malicious perturbations. This approach trains a neural network to predict the most effective perturbation for compromising upcoming speech signals. By requiring only a single computation to generate the perturbation, VoiceCamo enables real-time attacks targeting live speech, which results in a reduced recognition accuracy for ASR systems. However, it is important to note that these real-time malicious perturbation attacks are primarily focused on white-box models.

\section{Defense Approaches}
To protect against AI-driven attacks targeting VCS, various defensive strategies have been proposed.
We list the representative defense methods in Table~\ref{Tab: defense}.
\begin{table*}
\centering
\caption{Existing defense mechanisms against AI-driven attacks towards VCS.}
\label{Tab: defense}
\begin{tabular}{|c|c|c|c|c|c|c|c|} 
\hline
\multirow{2}{*}{Type}                                                         & \multirow{2}{*}{Defense Approach} & \multirow{2}{*}{Method} & \multicolumn{5}{c|}{Effectiveness}                                                                                                                                                                                                                                                                                          \\ 
\cline{4-8}
 &           &       & \begin{tabular}[c]{@{}c@{}}Deepfake \\Voice\end{tabular} & \begin{tabular}[c]{@{}c@{}}Backdoor \\Attack\end{tabular} & \begin{tabular}[c]{@{}c@{}}Unintelligible \\Speech\end{tabular} & \begin{tabular}[c]{@{}c@{}}Adversarial \\Example\end{tabular} & \begin{tabular}[c]{@{}c@{}}Malicious \\Perturbation\end{tabular}  \\ 
\hline
\multirow{4}{*}{\begin{tabular}[c]{@{}c@{}}Liveness~\\Detection\end{tabular}} & Void~\cite{ahmed2020void}                              & Spectrum           &  \CIRCLE                                                        &  \CIRCLE                                                         & \CIRCLE                                                                &  \CIRCLE                                                             &   \CIRCLE                                                               \\ 

\cline{2-8}
 & VoiceGesture~\cite{10.1145/3133956.3133962}         & Consistancy             &  \CIRCLE                                                        &  \Circle                                                         & \CIRCLE                                                                &  \CIRCLE                                                             &   \Circle                                                                   \\ 
 \cline{2-8}
 & Speaker-Sonar~\cite{10.1145/3380991}         & Consistancy             &  \CIRCLE                                                        &  \Circle                                                         & \CIRCLE                                                                &  \CIRCLE                                                             &   \Circle                                                                   \\ 
\cline{2-8}
      & CaField ~\cite{10.1145/3319535.3354248}                 & Fingerprinting          & \CIRCLE                                                        &  \Circle                                                         & \CIRCLE                                                                &  \CIRCLE                                                             &   \Circle                                                                   \\ 
\cline{2-8}
      & ARRAYID ~\cite{277088}                 & Fingerprinting          & \CIRCLE                                                        &  \Circle                                                         & \CIRCLE                                                                &  \CIRCLE                                                             &   \Circle                                                                   \\ 
\hline
\multirow{2}{*}{Purification}                                             & WaveGuard~ \cite{hussain2021waveguard}                        & Signal processing       &   \Circle                                                       & \Circle                                                          &    \Circle                                                             &   \CIRCLE                                                            &  \Circle                         \\ 
\cline{2-8}
& AudioPure   ~\cite{wu2022defending}                    & Diffusion model         &   \Circle                                                       & \Circle                                                          &    \Circle                                                             &   \CIRCLE                                                            &  \RIGHTcircle                                                                 \\ 

\hline
\multirow{2}{*}{\begin{tabular}[c]{@{}c@{}}Adversarial\\Example\end{tabular}} & Attack-VC~\cite{huang2021defending}                         & GD                      &   \CIRCLE                                                       &     \Circle                                                      &       \Circle                                                          &   \Circle                                                            &  \Circle                                                                 \\ 
\cline{2-8}
& VSMask~\cite{wang2023vsmask}                            & Predictive model        &   \CIRCLE                                                       &     \Circle                                                      &       \Circle                                                          &   \Circle                                                            &  \Circle                                            \\ 
\hline
\begin{tabular}[c]{@{}c@{}}Adversarial\\Training\end{tabular}                 & OCCAM~\cite{zheng2021black}                             & Adversarial training    &   \Circle                                                      &     \Circle                                                      &       \Circle                                                          &   \CIRCLE                                                            &  \RIGHTcircle                                                                 \\
\hline
\end{tabular}

\begin{tablenotes}\footnotesize
\centering
\item Effectiveness: \Circle:~not effective, \CIRCLE:~effective, \RIGHTcircle:~partically effective.
\end{tablenotes}
\end{table*}
\vspace{-10pt}
\subsection{Liveness detection}
All malicious audio signals are synthetically generated.
As a result, their playback has to be transmitted by loudspeakers rather than human mouth.
This characteristic makes it possible to distinguish between natural human speech and speaker-emitted sounds. 
Recent research underscores that natural human speech displays distinct spectral patterns compared to replayed audio, enabling sophisticated deep learning models to identify malicious audio with precision.
VoiceGesture~\cite{10.1145/3133956.3133962} and Speaker-Sonar~\cite{10.1145/3380991} introduce advanced methods that harness the speakers of mobile devices to emit ultrasounds, capturing unique gestures as users speak. These identifiable patterns are instrumental for two-factor authentication and pinpointing spoofed audio signals. The advent of multi-microphone-equipped mobile devices paves the way for extracting intricate patterns from multi-channel audio signals. Fieldprint~\cite{yan2019catcher} leverages dual-channel audio to craft a distinctive fingerprint aligned with the user’s identity. ARRAYID~\cite{277088} further enhances liveness detection by employing the integrated circular microphone array to derive speech fingerprints with enhanced accuracy and reliability.

\subsection{Purification}
However, the current liveness detection methods are exclusively designed to identify malicious audio signals, leaving benign voice commands exposed and prone to disruption. 
To bridge this gap, recent studies have unveiled audio signal purification as a viable countermeasure. 
In most adversarial audio instances, the power of the perturbation is notably low, rendering the malicious signal vulnerable to a range of signal transformation techniques. 
WaveGuard~\cite{hussain2021waveguard} incorporates various signal transformation strategies such as down-sampling, quantization, filtering, mel-spectrogram extraction, and linear predictive coding (LPC) to counter adversarial examples.
For benign audio signals, the inference outcomes are consistent pre and post-transformation. In contrast, adversarial examples witness a marked increase in the character error rate (CER), and these transformation techniques can, to an extent, recover the original benign speech content. 
Meanwhile, diffusion models are emerging and gaining traction for their strong generative abilities. For example, AudioPure~\cite{wu2022defending} introduces the incorporation of mild perturbations into adversarial examples. It employs reverse sampling to purify the audio and outputs cleared audio without adversarial noise.

\subsection{Defense Using Adversarial Perturbations}
While liveness detection defenses can identify deepfake voice attacks, their preventive measures are not as robust.
Attack-VC~\cite{huang2021defending} addresses this issue by implementing a defense strategy that employs adversarial examples. 
This method modifies perturbations based on the gradient of the speaker encoder in voice synthesis models, preventing the model from replicating the victim’s voice effectively.
However, its dependency on offline PGD optimization limits its applicability in real-time scenarios such as online meetings or voice instant messages. 
VSMask~\cite{wang2023vsmask} addresses this limitation using a predictive model-based defense mechanism that enables instantaneous perturbation generation, eliminating additional processing latency. 
In this way, it facilitates real-time speech protection against unauthorized voice synthesis attacks.

\subsection{Adversarial Machine Learning}
Adversarial machine learning has been considered as an effective method to reduce the impact of adversarial image examples~\cite{madry2018towards}. 
Similarly, this technique can be applied to detect raw audio content within adversarial examples in the audio domain.
OCCAM~\cite{zheng2021black} assesses the efficacy of adversarial training against these adversarial audio examples. Experimental results indicate a decline in the success rate of adversarial examples when tested against an ASR model that has been optimized using adversarial training. 
However, this enhanced robustness also reduces the recognition accuracy when processing clean audio. Additionally, adversarial training requires a significant amount of additional time ($>$10$\times$) compared to the standard training process.

\section{Evaluation}
\subsection{Evaluation Setup}
\begin{figure}
    \setlength{\abovecaptionskip}{0pt} 
    \centering
    \includegraphics[width=7.5cm]{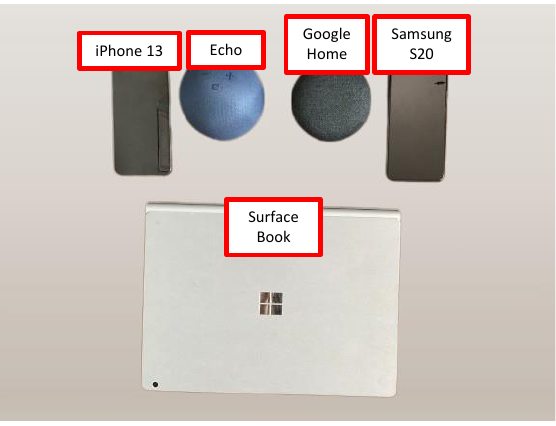}
    \caption{Devices for physical measurement studies.}
    \label{fig:devices}
\end{figure}

In this section, we present our evaluation results of different attacks in both physical and digital spaces.
For digital attack evaluation, we evaluate different attack methods on different commercial speech-to-text APIs, including Google Cloud, Amazon AWS, and Microsoft Azure.
Meanwhile, we measure the performance of  physical attacks on commercial voice assistant products, including Siri, Amazon Alexa, Google assistant, Samsung Bixby and Microsoft Cortana, as shown in Fig.~\ref{fig:devices}.
We use a JBL Charge 4 loudspeaker to play the audios in physical attack experiments. 
\subsection{Deepfake Voice Attacks}\label{deepfake evaluation}
We first evaluate the effectiveness of deepfake technology in compromising SV models. 
For this assessment, we ensure that all deepfake voices are transmitted over-the-air. 
As highlighted in Section~\ref{deepfakeattack}, AdaIN-VC and AutoVC are two VC models capable of synthesizing speech using a short speech sample from the target speaker. 
We select  two commercial voice cloning services,  resemble.ai~\cite{resembleai} and play.ht~\cite{playht}, in this evaluation.  Thse services offer two distinct voice cloning modes. The instant voice cloning option (I) allows users to upload a short speech sample, and the service employs a pre-trained model to generate the synthetic voice. 
In contrast, the high-quality synthesis mode (H) requires multiple speech samples that are incorporated into the training dataset to refine the synthetic voice’s quality for the given speaker.

In this evaluation, we collect speech samples from 10 volunteers, using them to generate synthetic voice commands intended to deceive voice assistants. To assess the similarity between these deepfake voices and the original samples, we utilize SpeechBrain~\cite{speechbrain}, a leading-edge SV model. SpeechBrain is notable for its implementation of the Time Delay Neural Network (ECAPA-TDNN) framework, which incorporates emphasized channel attention, propagation, and aggregation within the network. By employing SpeechBrain, we extract unique voice embeddings, enabling an in-depth comparison of the synthetic and authentic voices of each participant. In this study, we set the default verification threshold of SpeechBrain at 0.25. This setting is chosen for its optimal performance in experiments involving clean speech datasets.

\begin{figure*}[tbp]
    \setlength{\abovecaptionskip}{0pt} 
    \centering
    \subfigure[The similarity scores of deepfake voice attacks.]{\includegraphics[height=3.7cm]{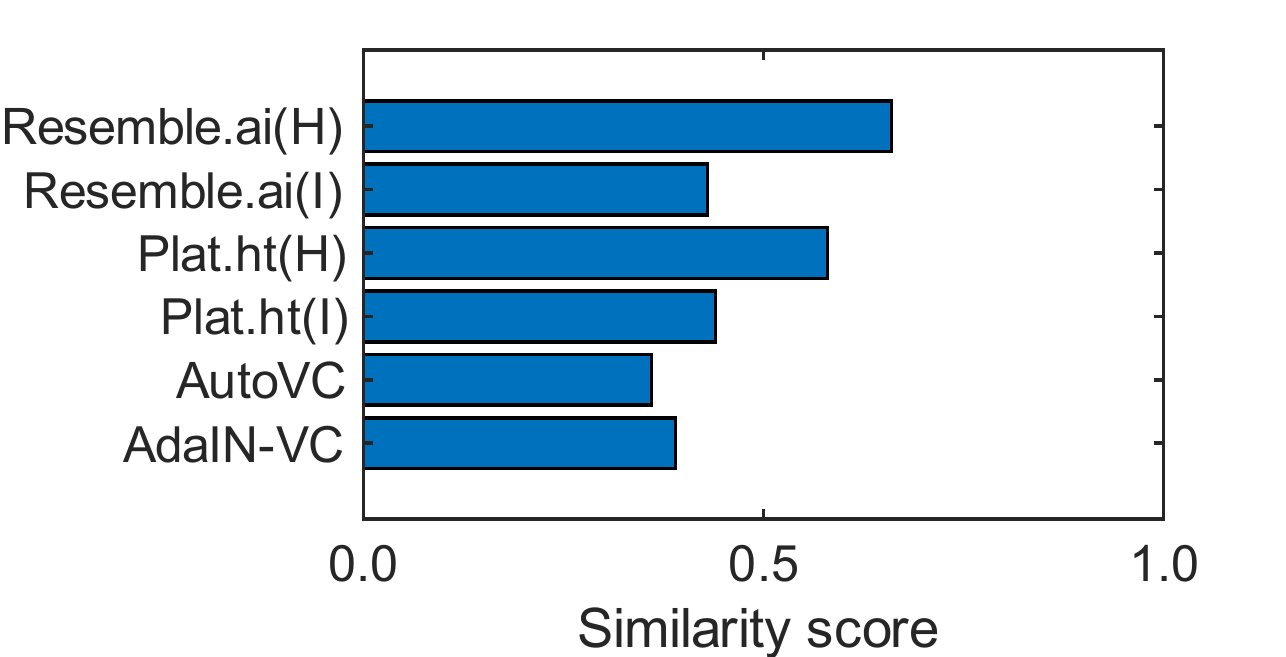}\label{fig:score}}
    \
    \subfigure[The attack success rate of deepfake voice attacks.]{\includegraphics[height=3.7cm]{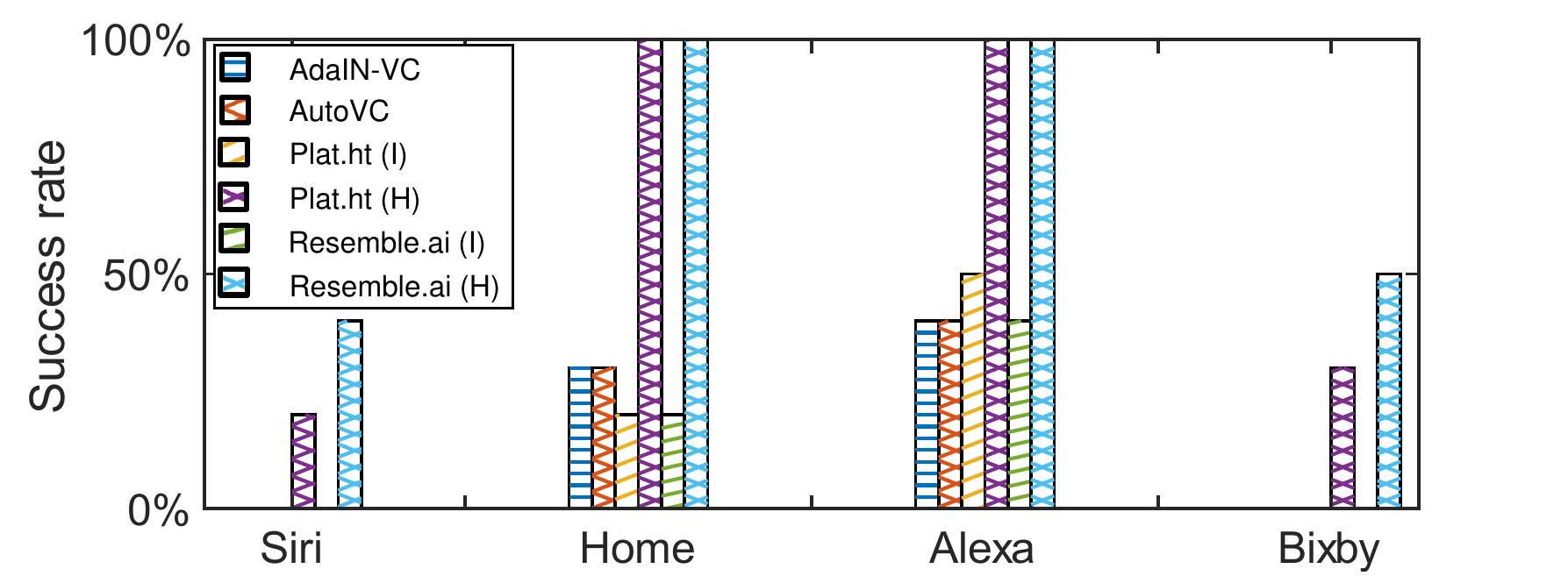}\label{fig: rates}}
    \caption{The attack performance of different deepfake voice generation methods.}
    \label{fig: deep_fake_performance}
\end{figure*}

The results of our experiments are shown in Fig.~\ref{fig: deep_fake_performance}.
While SpeechBrain provides text-independent speaker verification, it is not as resilient as text-dependent verification methods.
All the deepfake voice generation methods outlined in this study have successfully bypassed the SpeechBrain SV model.

In contrast, commercial VCS display exceptional robustness when confronted with deepfake voice attacks.
Based on our experimental findings, mobile devices with text-dependent SV, such as Apple Siri and Samsung Bixby, exhibit the highest level of robustness against deepfake voice attacks. We utilize various voice synthesis models to create wake-up words and attempt to activate the voice assistant without unlocking the smartphone.
For VC models and commercial text-to-speech APIs that employ the ``instant voice cloning" mode, None of synthetic speech examples manage to activate the voice assistant of Siri and Bixby using their owners' voice.

High-fidelity voice cloning models utilize an extensive amount of training data to optimize a  pre-trained model for individual speakers, which outperform VC and instant voice cloning models. 
In SpeechBrain, these models achieve a voice similarity score exceeding 0.5, which is sufficient to bypass both text-independent SV models and human perception.
However, Apple Siri and Bixby are still robust against these sophisticated attackers.
With the most powerful deepfake voice generation method, 50\% or more samples fail to fool their SV processes.
One potential reason is that  text-dependent speaker verification not only learns the voice itself, but also it  extracts the speaking style of the verified user.
Consequently, both VC and TTS synthesis methods are inadequate to generate qualified wake-up words for all target speakers.
Consider that high-fidelity voice cloning requires a large-scale training data, e.g., more than 20 speech samples with a given content, it is non-trivial for the adversaries to launch these attacks.
Meanwhile, Amazon Alexa does not have a speaker verification process, meaning that anyone speakers ``Alexa" to activate the voice assistant.
The only case where Alexa requires speaker verification is for some privacy-related operations, e.g., emails or online purchases.
As a result, it is easier for the attackers to activate Alexa and initiate malicious commands.

\subsection{Fake Wake-up Word Attacks}

We evaluate the effectiveness of fake wake-up words, as shown in Table ~\ref{Tab: fake-wake}. 
In this measurement evaluation, fake wake-up words are produced using genetic algorithm method, and a standard male voice is used to turn these ambiguous words into audible speech. 
The evaluation results show that all tested VCS devices are susceptible to specific fake wake-up words similar to the real wake-up words.
Among all devices, Amazon Alexa is the most vulnerable VCS.
 This is primarily because its wake-up word, ``Alexa", is short, thereby increasing the likelihood of identifying similarly sounding words that can inadvertently activate the assistant. 
Google Home and Apple Siri, whose wake-up commands consist of two words, exhibit better resilience against such attacks.
It is important to note that the successful fake wake-up words are audibly similar to the actual ones, to the extent that human listeners might also recognize them as real wake-up words. For example, ``hey sirr e" is perceived as close to ``Hey Siri", differentiated mainly by pauses. Furthermore, as mentioned in Section~\ref{deepfake evaluation}, both Apple and Google devices are sensitive to human speaking styles in real-world scenarios, making it challenging to deceive the SV process with synthesized fake wake-up words, especially for those using text-dependent SV models.

\begin{table}
\centering
\caption{Most effective fake wake-up word samples and their success rates.}
\label{Tab: fake-wake}
\begin{tabular}{|c|c|c|c|c|} 
\hline
                                               & Echo                                      & Google Home                               & Apple Siri                                & Bixby                                     \\ 
\hline
Wake-up word                                   & Alexa                                     & Hey Google                                & Hey Siri                                  & Hi Bixby                                  \\ 
\hline
\hline
\multirow{2}{*}{}                              & ileqsar                                   & hey gooogov a                             & hey sirr e                                & hai bikisibi                              \\ 
\hhline{|~----|}
                                               & {\cellcolor[rgb]{0.753,0.753,0.753}}100\% & {\cellcolor[rgb]{0.753,0.753,0.753}}100\% & {\cellcolor[rgb]{0.753,0.753,0.753}}100\% & {\cellcolor[rgb]{0.753,0.753,0.753}}80\%  \\ 
\hline
Top-3 samples                                  & ilexsar                                   & hei googll a                              & hay syrrie e                              & hai bikisbee                              \\ 
\hline
\rowcolor[rgb]{0.753,0.753,0.753} Wake-up rate & 100\%                                     & 100\%                                     & 100\%                                     & 70\%                                      \\ 
\hline
\multirow{2}{*}{}                              & ilexsur                                   & heay gugal                                & hey sserea                                & hai biksi B                               \\ 
\hhline{|~----|}
                                               & {\cellcolor[rgb]{0.753,0.753,0.753}}80\%  & {\cellcolor[rgb]{0.753,0.753,0.753}}100\% & {\cellcolor[rgb]{0.753,0.753,0.753}}80\%  & {\cellcolor[rgb]{0.753,0.753,0.753}}70\%  \\
\hline
\end{tabular}
\end{table}

\subsection{Unintelligible Speech}
In the past, unintelligible speech attacks have been tested and shown effectiveness against commercial VCS. These attacks involve transforming adversarial audio examples using basic signal processing methods, eliminating the need for detailed knowledge about the victim's speech recognition models.  However, as commercial VCS products and their underlying models continuously evolve, it is essential to re-examine the effectiveness of these attacks.

In our evaluation, we replicate two typical unintelligible speech attacks: ``hidden commands"~\cite{carlini2016hidden} and ``practical hidden voice"~\cite{abdullah2019practical}. We test their performance in both digital and over-the-air scenarios. For the digital attacks, audio samples are directly uploaded to the speech-to-text APIs. In the over-the-air attack, a JBL loudspeaker serves as the attack device, with the microphone placed 50 cm away and the volume set around 75 dBA.

We list the measurement results in Table~\ref{Tab: unintell}. First, we measure the efficacy of ``hidden commands" on 5 different models or products.
We observe enhanced robustness in current ASR models against unintelligible speech attacks. 
None of the unintelligible wake-up pharse, such as ``Alexa" or ``Ok Google", are accurately recognized by ASR models. 
Some samples, for instance, ``tweet goodbye", are merely interpreted as ``goodbye". 
Nonetheless, the models demonstrate high accuracy in recognizing longer phrases like ``turn on airplane mode", owing to their ability to contextualize and infer the complete command.
Interestingly, these supposed unintelligible voice commands retain a degree of intelligibility to the human ear. We employ the Short-Time Objective Intelligibility measure (STOI) to evaluate this aspect~\cite{taal2011algorithm}. The average STOI of these unintelligible speeches remains above 0.5, indicating a potential for human listeners to discern the embedded malicious commands.

For ``practical hidden voice" attack, we apply the same experimental setup as hidden commands. For the phase shift technique, we randomly alter 50\% of phases in the audio signal. We also employ time-inversion, and apply a 5 ms time window, and time scaling, adjusting to 50\% of the original signal’s duration. The continual enhancement of ASR models has resulted in a noticeable decline in attack success. Random phase-shifted unintelligible speech is unrecognizable across all tested models. While online APIs maintain high accuracy in interpreting unintelligible speech generated through time inversion or time scaling, their performance dips significantly when both signal processing methods are combined. In such cases, Google and Amazon APIs fail to interpret the modified audio, while Microsoft API manages to transcribe some samples with reduced precision.
The attack success rate further decreases when the unintelligible speech samples are injected through over-the-air transmission. In over-the-air attacks, 
Google Home, Amazon Echo, and Cortana only recognize a few unintelligible speeches. Siri and Bixby stand out as the most robust voice command systems, with none of our samples being able to manipulate them. This resistance can be attributed to two factors. First, the transmission loss during airborne transmission degrades the audio quality, reducing the effectiveness of the attack. Second, the ASR models embedded in these voice assistants are designed to be less sensitive to noisy audio signals to avoid accidental activation, further impeding the effectiveness of unintelligible speech attacks. 

\begin{table*}
\centering
\caption{Unintelligible speech attack performance in real world scenarios.}
\label{Tab: unintell}
\begin{tabular}{|c|c|l|c|c|c|c|c|c|c|c|} 
\hline
Attack                                                   & Method                                                       & STOI & Google API & AWS API & Azure API & Google Home & Echo & Siri & Bixby & Cortana  \\ 
\hline
\begin{tabular}[c]{@{}c@{}}Hidden \end{tabular} & \begin{tabular}[c]{@{}c@{}}MFCC Recon.\end{tabular} &  0.57    & 9/10       & 9/10       & 10/10  & 1/10        & 2/10 & 0/10 & 0/10  & 1/10     \\ 
\hline
\multirow{4}{*}{Practical}                               & Phase Shift~                                                 & 0.06     & 0/10       & 0/10       & 0/10   & 0/10        & 0/10 & 0/10 & 0/10  & 0/10     \\ 
\cline{2-11}
                                                         & Time Inversion~                                              & 0.42     & 9/10       & 8/10       & 10/10  & 2/10        & 3/10 & 0/10 & 0/10  & 2/10      \\ 
\cline{2-11}
                                                         & Time Scaling~                                                &  N/A    & 5/10       & 5/10       & 7/10   & 1/10        & 1/10 & 0/10 & 0/10  & 2/10     \\ 
\cline{2-11}
                                                         & Inversion~+~Scaling                                            &  N/A    & 0/10       & 0/10       & 6/10   & 0/10        & 0/10 & 0/10 & 0/10  & 0/10     \\
\hline
\end{tabular}
\end{table*}

\subsection{Adversarial and Backdoor Attacks Targeting SV}
In this section, we conduct measurement experiments on various commercial VCS to evaluate adversarial and backdoor attacks targeting SV models. 
Initially, we test the performance of FakeBob attack~\cite{chen2021real}. 
Adversarial examples were generated using both i-vector and GMM models.
It is worth noting that all commercial VCS employ SV models that only provide binary output labels such as ``yes" or ``no". Consequently, the success of attacks would depend on the transferability of the adversarial examples. 

First, we use digital injection method to attack the original target SV models.
With $\sim$20 dB SNR, the adversarial examples can pass the SV models with 100\% success rate.
Next, we increase the perturbation power until SNR reaches $\sim$0 dB and launch the attack over-the-air.
On SpeechBrain~\cite{speechbrain} model, none of the samples targeting GMM or i-vector succeeds as SpeechBrain uses ECAPA-TDNN to extract voice patterns.
Finally, we test the adversarial samples on commercial SV systems.
Surprisingly, all the SV platforms are robust to the adversarial attack.
There are two potential reasons for the robustness.
First, the target models (i-vector \& GMM) and SpeechBrain are light-weighted SV models.
In comparison, commercial SV platforms are based on large-scale models, and they work under text-dependent mode, for which the AEs show limited transferability.
Second, commercial VCS might refuse noisy wake-up voice commands to avoid mis-activation.

\begin{figure}[tbp]
    \setlength{\abovecaptionskip}{0pt} 
    \centering
    \includegraphics[width=7cm]{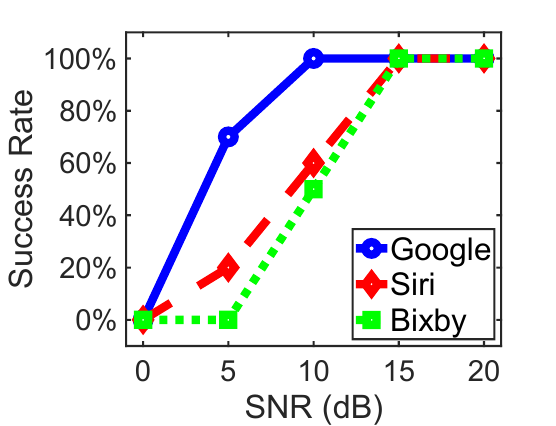}
    \caption{Wake-up success rate of replay voice commands with different noise levels.}
    \label{fig:noise_levels}
\end{figure}

To validate this conjecture, we play white noise with different volumes and replay registered wake-up commands to three different VCS including Google Home, Siri, and Bixby.
Fig.~\ref{fig:noise_levels} shows the evaluation result.
While the noise level is low ($\geq 15$ dB), all mobile devices can accurately recognize the voice and wake up the voice assistant.
For Siri and Bixby, they both have lower sensitivity in noisy environments.
When the SNR reaches 5 dB, only 20\% wake-up commands can be recognized by Siri, and no commands can be recognized by Bixby.
For Google Assistant, it is more robust in noisy environments.
It can still recognize 70\% wake-up commands when SNR reaches 5 dB.
However, as Google Assistant is a text-dependant and large-scale SV model, current adversarial attacks could not inject malicious comments.
For FakeBob attack, the SNR of effective AEs is $\sim 0$ dB.
This level makes it challenging for VCS to recognize the input, even if it successfully deceives the SV model.

For backdoor attacks, such as FenceSitter\cite{deng2022fencesitter}, the primary target is to compromise text-independent SV models. The attackers inject a backdoor into the enrolled speech samples and subsequently utilize a trigger audio to bypass the SV model. However, this approach encounters a significant obstacle; the trigger audio is not recognized as the wake-up word by commercial VCS, which predominantly rely on text-dependent SV models.
Consequently, the backdoor attacks will be invalid for commercial text-dependent SV models.
Currently, the compromise of commercial text-dependent SV models remains an underdeveloped area. This presents a potential direction for future research in VCS security.

\begin{table}
\centering
\caption{AE's attack performance against  SV systems.}
\label{Tab: AE_SV}
\begin{tabular}{|c|c|c|c|c|c|} 
\hline
        & Target & SpeechBrain   & Google & Siri & Bixby   \\ 
\hline
i-vector & 10/10  & 0/10 & 0/10   & 0/10 & 0/10      \\ 
\hline
GMM     & 10/10  & 0/10 & 0/10   & 0/10 & 0/10     \\
\hline
\end{tabular}
\end{table}

\subsection{Adversarial Attacks Targeting ASR}
We select representative adversarial attacks that allow airborne transmission for our evaluation.  Table~\ref{Tab: AE_ASR} shows the attack performance. 
These attacks are evaluated on both VCS APIs and voice assistant products. In the case of over-the-air attacks, we only count successful cases that adversarial examples effectively trigger the malicious commands of VCS. 

The ``Robust"~\cite{carlini2018audio} attack, which specifically targets the DeepSpeech model, demonstrates limited effectiveness on other commercial VCS. In the context of digital attacks, only a few words get recognized by ASR. 
Notably, for over-the-air attacks, despite the substantial perturbation power ($\sim$0 dB), all attempts to compromise ASRs fail.

Next, we proceed to test the ``Devil's Whisper" attack~\cite{chen2020devil}.
We generate AEs using gradient estimation and inject them via digital methods.
In this scenario, the AEs exhibit impressive performance. On Google and Azure speech-to-text APIs, most AEs are accurately recognized by ASR models and interpreted as the corresponding malicious voice commands. 
However, ``Devil's Whisper" fall short in compromising commercial VCS through airborne transmission. 
For Siri and Bixby, as there are no open APIs for gradient estimation, we utilize various AEs from other target models, but none manages to fool them. 
Even for Google Home, Alexa, and Cortana, which have corresponding API services, the AEs could not succeed. 
To understand the cause of this failure, we embed the distinct command within the carrier music, maintaining the same signal power, and replayed them through over-the-air transmission. In contrast, ASR models recognized all these audio samples, suggesting that the carrier music does not obscure the malicious command. However, for commercial devices, pre-processing might eliminate the hidden perturbation, rendering them undetectable to VCS.

In addition, we also test some real-time attacks via malicious perturbation.
As real-time perturbations only work in a short-time, it requires the knoweledge of white-box models to compromise the benign voice commands.
According to our measurement results, the real-time malicious perturbations show poor transferability for black-box models.
For both digital and over-the-air attacks, the ASR models can always identify the true transcriptions. 
As a result, current malicious perturbation attacks cannot successfully compromise commercial ASR models especially when considering human-in-the-loop conditions.

Lastly, we evaluate SMACK~\cite{yusmack}, a semantically meaningful attack. Unlike other attacks, SMACK retains intelligibility and clarity, leading to better attack performance compared to traditional adversarial attacks. 
We test SMACK samples specifically targeting the Google speech-to-text API. Our findings indicate that 8 out of 10 AEs successfully compromise the ASR model on the Google API. Moreover, in the case of over-the-air attacks, 4 out of 10 still prove effective on Google Home. 
SMACK demonstrates transferability in attacking other commercial VCS. Certain AEs are effective against Siri without necessitating phoneme adjustments. 
Some SMACK AEs can also temporarily compromise Samsung Bixby. However, Bixby possesses the capability to automatically correct certain illogical transcriptions. For instance, an AE ``typhoon is coming" transcribed as ``an iPhone is coming" will be rectified by Bixby, discarding the nonsensical interpretation and opting for a logical transcription from phonemes that sound similar.
However, SMACK is limited to generating malicious audio that closely resembles benign speech carriers, implying that these sounds are perceptible to human ears.
\vspace{-10pt}



\begin{table*}
\centering
\caption{AE's attack performance for commercial ASR models of VCS.}
\label{Tab: AE_ASR}
\begin{tabular}{|c|c|c|c|c|c|c|c|c|} 
\hline
Attack Method & \multicolumn{3}{c|}{Digital Attack} & \multicolumn{5}{c|}{Over-the-air Attack}  \\ 
\hline
Target Model             & AWS   & Google & Azure & Siri & Home & Echo & Bixby & Cortana      \\ 
\hline
Robust            & 0/10 & 0/10   & 0/10                    & 0/10 & 0/10 & 0/10 & 0/10  & 0/10         \\ 
\hline
Devil             & 2/10 & 10/10  & 10/10                    & 0/10 & 0/10 & 0/10 & 0/10  & 0/10         \\ 
\hline
Spec              & 0/10      &  0/10      &   0/10                   & 0/10    &0/10    &0/10   &0/10    &0/10              \\ 
\hline
SMACK             &     0/10  &     8/10   &  1/10                       &   3/10   &  4/10    &   0/10   &  0/10     &  1/10            \\
\hline
\end{tabular}
\end{table*}

\subsection{Defense Effectiveness}
In this section, we measure various defense mechanisms to evaluate their effectiveness against existing attacks. We select Void~\cite{ahmed2020void} and WaveGuard~\cite{hussain2021waveguard} as representative defenses that are based on audio patterns and signal processing, respectively. We use Resemble.ai high-quality mode to generate deepfake voice samples.

Meanwhile, we use an iPhone 13 Pro to capture live human speech along with malicious audio samples that are played back through a loudspeaker. The performance metrics of Void are detailed in Table~\ref{Tab: void_performance}.
\begin{table}
\centering
\caption{Void defense performance against different attack methods.}
\label{Tab: void_performance}
\begin{tabular}{|c|c|c|c|c|c|c|c|} 
\hline
Attack    & None & Replay & DV & FB & SP & DW & SMACK  \\ 
\hline
Pass Rate & 77\% & 26\%   & 18\%      & 0\%     & 0\%   & 0\%   & 20\%    \\
\hline
\end{tabular}
\begin{tablenotes}\footnotesize
\centering
\item DV: Deepfake Voice; FB: FakeBob; SP: SpecPatch; DW: Devil's Whisper.
\end{tablenotes}
\end{table}
From the measurement result, we find that Void can identify more than 70\% simple replayed voice command samples.
For deepfake voice attack and adversarial attacks, Void achieves very high performance.
Only 10\% deepfake voice generated by Resemble.ai can compromise the Void model.
Other adverarial examples could not bypass Void, as their perturbations exhibit abnormal patterns in the audio features. 
 Interestingly, some live human speech also could not pass through Void. This high false rejection rate could impact the sensitivity of VCS.

Next, we apply WaveGuard to process the adversarial examples. During the evaluation, attacks are initiated on the original target models. We count a successful defense if the purified audio is identified as a different identity or transcription. 
We explore various signal processing defense methods offered by WaveGuard.

In the scenario of the digital FakeBob attack, the effectiveness varies among the different defense methods. Notably, downsampling-upsampling and mel-spectrogram reconstruction emerge as the most proficient in countering adversarial examples. In contrast, for airborne attacks, LPC and Quantization are no longer effective due to the excessive strength of the perturbations.
Despite this, downsampling-upsampling and mel-spectrogram reconstruction maintain their effectiveness against the majority of attacks but at the expense of significant audio quality degradation. To bolster the identification of AI-driven attacks, integrating liveness detection with signal processing proves beneficial, resulting in enhanced accuracy in distinguishing malicious intrusions.

\begin{table}
\centering
\caption{WaveGuard defense performance against different attack methods.}
\label{Tab: AE_waveguard_performance}
\begin{tabular}{|c|c|c|c|c|} 
\hline
\diagbox{Attack}{Method} & Down. & Quan. & Mel. & LPC  \\ 
\hline
FakeBob (Digital)              &   10/10    & 5/10      & 10/10     & 6/10     \\ 
\hline
FakeBob (Airborne)              &   7/10    & 1/10      & 5/10     & 0/10     \\ 
\hline
Robust                   &  6/10     & 0/10      & 5/10     & 0/10     \\ 
\hline
Devil's Whisper                    &   8/10    & 2/10      & 9/10     & 2/10     \\ 
\hline
SpecPatch                &   10/10    & 2/10      & 9/10     &  0/10    \\ 
\hline
SMACK                    &  3/10     & 0/10      & 1/10     &  0/10    \\
\hline
\end{tabular}
\end{table}

\section{Insights}
Based on our evaluation, current attack methodologies do not achieve the expected attack effectiveness.
VCS appear to be more robust against a range of attacks than initially anticipated. 
These systems have demonstrated a heightened level of security, successfully countering various intrusion attempts. 
In this section, we summarize the key insights in the effectiveness of both prevailing attack strategies and the defense mechanisms in place. 
The resilience exhibited by VCS highlights the advancements in their design and the imperative for ongoing enhancements to counter evolving threats.

\vspace{-5pt}
\subsection{Insights for Attacks}
\noindent \textbf{Transferability.} The results of our measurements indicate that only deepfake voices display attack transferability, meaning they can attack different VCS indiscriminately with the same generative model. For the majority of backdoor and adversarial attacks, a large number of queries are required for gradient estimation. 
However, commercial SV models only output hard labels such as ``accept" or ``reject", making it difficult to optimize the adversarial examples. 
Moreover, the adversarial examples for ASR possess limited transferability.
Only a few samples that are optimized with white-box models manage to be accurately transcribed by commercial ASR platforms. 
Considering that attackers might not always know the specific VCS in use, there is a necessity for attacks with enhanced transferability across different products.
\\
\noindent \textbf{Perturbation Strength.} While increased perturbation strength can enhance the efficacy of adversarial attacks, particularly in over-the-air transmission, it often proves ineffective in real-world scenarios. 
This is because voice assistants are designed to reject commands delivered in noisy environments to ensure accuracy and security. 
For instance, FakeBob is effective when the SDR is approximately 0 dB.
However, in these instances, voice commands are too noisy to be recognized and processed by commercial VCS, which consequently fortifies the systems' robustness against such attacks.\\
\noindent \textbf{Human Factors.} Certainly, the human factor is essential for attackers, as user awareness of malicious voice commands can impact the attack’s effectiveness. 
Recent research has taken human factors into account, by designing attack mechanisms such as a short general ``pulse" perturbation~\cite{li2020advpulse} or a spectrogram ``patch"~\cite{guo2022specpatch} to mute the following command, aiming to minimize the interference caused by human activities.
However, these attacks exhibit weak transferability on commercial platforms. 
Compromising commercial ASR while considering human factors remains a significant challenge.
\vspace{-10pt}

\subsection{Insights for Defenses}
\noindent \textbf{Trade-off between Sensitivity and Robustness.}
We test various attacks on multiple commercial products and observe that Apple Siri and Samsung Bixby exhibit remarkable resilience to both deepfake voice and adversarial attacks. 
However, they also display the lowest sensitivity. In noisy environments or when users are at a distance from the device, it can be difficult to wake up these voice assistants and initiate further actions. 
In contrast, Alexa demonstrates the highest sensitivity, attributed to the absence of a text-dependent SV model for verifying the speaker’s identity. 
This, however, makes it susceptible to attacks that manipulate it to execute malicious commands. Nonetheless, since Alexa is typically stationed in private spaces, remotely delivering malicious comments without physical access presents a significant challenge for attackers.\\
\noindent \textbf{Defense Effectiveness.}
Based on the measurement results of existing defense methods, liveness detection models are adept at identifying most malicious audios. 
However, lightweight liveness detection models might also face a high false rejection rate. Other signal processing methods can reduce the effect of malicious audio signals and impede the performance of standard adversarial attacks. 
But when the perturbation is intense, as in an over-the-air attack, such signal processing solutions struggle to restore the original audio effectively. 
This is especially the case for semantically meaningful adversarial examples, where signal processing does not impact the transcription.
\vspace{-10pt}

\section{Conclusion}

A variety of AI-driven malicious audio attacks pose a threat to VCS. However, as these threats evolve, VCS are also continuously improving their pre-processing techniques and AI models. Our evaluation of the robustness of various commercial products against these advanced attacks demonstrates that these systems are more resilient than initially anticipated. Despite recognizing their robustness, our analysis also exposes specific vulnerabilities inherent in modern VCS, such as susceptibility to deepfake voice attacks, disruption by unintelligible speech, and manipulation through carefully crafted adversarial examples. Furthermore, we offer insights that are essential for the development of both attack and defense strategies, ensuring improved security in the rapidly evolving technological environment.
\bibliographystyle{IEEEtran}
\bibliography{reference}
\end{document}